\documentclass[twocolumn]{aa}  

\usepackage{graphicx}
\usepackage{txfonts}
\usepackage{lipsum}
\usepackage{subcaption}         
\usepackage{lscape}             
\usepackage{placeins}           

\usepackage[colorlinks=true, allcolors=blue]{hyperref}
\usepackage{orcidlink}

\begin{document} 

    \title{From active nests to fast wind gusts: Nested flux emergence as a precursor of persistent non-axisymmetric coronal holes}

   \titlerunning{From active nests to fast-wind gusts}
   \authorrunning{Finley et al.}

   \author{A. J. Finley~\orcidlink{0000-0002-3020-9409}\inst{1},
          D. M\"{u}ller~\orcidlink{0000-0001-9027-9954}\inst{1},
          M. Janvier~\orcidlink{0000-0002-6203-5239}\inst{1},
          M. Temmer~\orcidlink{0000-0003-4867-7558}\inst{2},
          Y. J. Rivera~\orcidlink{0000-0002-8748-2123}\inst{3},
          S. T. Badman~\orcidlink{0000-0002-6145-436X}\inst{3}
          }

    \institute{
    European Space Agency, ESTEC, Noordwijk, The Netherlands\\ \email{adam.finley.astro@gmail.com} 
    \and
    Institute of Physics, University of Graz, Austria
    \and 
    Center for Astrophysics, Harvard \& Smithsonian, Cambridge, Massachusetts, 02138, USA
    }

   \date{Received Oct 30, 2026; accepted -- --, 202-}

\abstract{Recurring flux emergence within active nests reshapes the Sun's large-scale magnetic field and creates unipolar magnetic regions that can support coronal holes. The fast solar wind emerging from coronal holes drives space weather across the heliosphere.}
{We investigated how the emergence of active nests at the peak of solar cycle 25 drove changes in global magnetic topology, coronal hole formation, and solar wind speed.}
{Coronal holes were identified from composite SDO/AIA Carrington maps and their magnetic field properties were extracted from SDO/HMI magnetograms. Spherical harmonic decomposition of the magnetograms quantified how photospheric magnetic flux was distributed across different spatial scales and was used to extrapolate the coronal magnetic field. We compared coronal structure with SOHO/LASCO-C2 coronagraph observations and the solar wind measured at 1~au.}
{Emerging magnetic flux within active nests rapidly surpassed the pre-existing photospheric flux in 2024, strengthening the non-axisymmetric magnetic field and shifting the corona into a tilted-dipole state. This dipolarisation stabilised the large-scale magnetic field, with the heliospheric current sheet anchored above active nests. Unipolar magnetic regions from the edges of active nests sustained non-axisymmetric coronal holes covering 6 to 14$\%$ of the solar surface in 2025, comparable to the $12\%$ covered by polar coronal holes at solar minimum. Flux emergence in both hemispheres near the Carrington longitude of the tilted-dipole's negative pole returned the corona to a quadrupole-dominated state in 2026.}
{Active nests helped stabilise the coronal magnetic field, while the decay and transport of unipolar magnetic flux sustained persistent, non-axisymmetric coronal holes. Differential rotation between low- and high-latitude unipolar regions triggered episodic expansion and contraction of the largest coronal holes. Early identification of specific active nest configurations, combined with solar cycle phase information, could improve the forecasting of coronal holes and fast solar wind streams.}

\keywords{Solar Magnetism -- Solar Wind -- Solar Activity --
                    }

\maketitle
\nolinenumbers

%

\section{Introduction}\label{sect:intro}

Dynamo-action within the convection zone drives the evolution of the Sun's magnetic field over an 11-year cycle \citep[for a review, see][]{brun2017magnetism}. Magnetic flux emerges at the photosphere, forming active regions that produce flares and coronal mass ejections \citep{van2015evolution}, whilst modifying the large-scale magnetic field topology \citep{reville2017global}. As the solar cycle progresses, active regions shift the distribution of photospheric magnetic field from low-degree, axisymmetric (aligned with the rotation axis) spherical harmonics toward more complex, non-axisymmetric components \citep{derosa2012solar, finley2023evolution}. Eventually, the transport and cancellation of emerged magnetic flux reverses the Sun's polar magnetic fields \citep{hathaway2015solar}. The evolution of the Sun's magnetic field shapes the heliosphere, organising both the fast and slow solar wind streams \citep{wang2010formation}. The fast solar wind presents a major space weather hazard, driving co-rotating interaction regions and preconditioning interplanetary space for the propagation of coronal mass ejections and energetic particle events \citep[for a review, see][]{richardson2018solar}.

Coronal holes, dark regions of the Sun's surface at extreme-ultraviolet (EUV) and soft x-ray wavelengths \citep{timothy1975structure, bohlin1977extreme, zirker1977coronal}, are the source of the fastest solar wind streams at 1~au (600 to 800 km/s). Coronal holes have lower temperatures and densities than their surroundings, and contain relatively weak magnetic field with a net magnetic polarity \citep[for a review, see][]{cranmer2009coronal}. The unipolar magnetic regions within coronal holes are comprised of small-scale funnels (at granular and super-granular scales) that open into the large-scale magnetic field and solar wind above \citep{wedemeyer2009coupling, finley2022stirring}. Polar coronal holes appear when meridional transport of magnetic flux from the previous cycle accumulates unipolar magnetic field at the Sun's poles \citep{sun2015polar}. During the solar cycle, the emergence and decay of active regions creates locally imbalanced magnetic flux that allows non-axisymmetric coronal holes to develop \citep{karachik2010formation, van2012magnetic, van2015evolution}. The rotation of non-axisymmetric coronal holes is enforced by the large-scale magnetic field, with interchange reconnection allowing their boundaries to slip relative to the surface differential rotation \citep{wang1988quasi, wang1993understanding}. The longitudinal drift of coronal holes over several months reflects their connection to flux emergence and solar cycle phase \citep{krista2018longitudinal, heinemann2020statistical, harris2022tracking}. Long-lived active region complexes, or ``active nests'', produce large unipolar magnetic regions that can support persistent coronal holes and affect the polar field reversal \citep{mordvinov2013decay, mordvinov2014reversals, golubeva2016decay}.

Active nests contain recurring flux emergence that powers eruptive activity over several solar rotations \citep{temmer2005causes, gyenge2016active, loumou2018association, korsos2025solar, finley2025prolific}. Nearly 50\% of all active regions are formed by nested flux emergence, appearing in close proximity to previously emerged flux \citep{castenmiller1986sunspot, harvey1993properties, schrijver2008solar, norton2025moderate}. Active nests tend to anchor the heliospheric current sheet (HCS) directly above them \citep{balthasar1983preferred, neugebauer2000solar, benevolenskaya2005formation, obridko2011active, yang2024observing, tahtinen2024straight}. They influence the Sun's large-scale magnetic field by amplifying non-axisymmetric magnetic field components \citep{tahtinen2026active, tahtinen2026amplification} and stalling the reversal of the coronal magnetic field \citep{finley2024nested}. In the later stages of solar cycle 24, nesting rejuvenated the dipole component of the Sun's magnetic field \citep{sheeley2015recent}, giving rise to large non-axisymmetric coronal holes \citep{yoshida2026temporal} that increased the open magnetic flux in the heliosphere \citep{heinemann2024origin} and contributed to the polar field strength at solar minimum \citep{wang2025solar}.  

At the maximum of solar cycle 25 in 2024 \citep{miesch2025solar, jouve2025forecasting}, active nests dominated solar activity and reinforced a tilted-dipole configuration in the coronal magnetic field \citep{finley2026active}. Parker Solar Probe \citep{fox2016solar} and Solar Orbiter \citep{muller2020solar} explored the inner heliosphere during this tilted-dipole state, assessing how the fast solar wind developed from alfvénic turbulence at ten solar radii \citep{bandyopadhyay2026first}, to a flow of $\sim800$~km/s at 1~au \citep{badman2026direct}. The heliospheric magnetic field in the ecliptic plane resembled Ulysses' fast latitude scan of the axisymmetric dipole at solar minimum \citep{mccomas2000solar}. However, rather than two broad fast wind streams originating from dipolar coronal holes, local sub-structure produced a fragmented outflow \citep[discussed in][]{rivera2026dipole}. 


This study investigates the formation of persistent non-axisymmetric coronal holes, linked to flux emergence in active nests and the stability of the coronal magnetic field. Section~\ref{sect:2} describes the data products and identification method for coronal holes. Section~\ref{sect:3} provides an overview of solar activity, flux emergence, large-scale magnetic field evolution, and coronal hole growth over solar cycle 25. We then highlight the largest continuous non-axisymmetric coronal hole, that persisted for almost two years. In Sect.~\ref{sect:4}, we discuss how nested flux emergence shaped the coronal magnetic field and produced the unipolar magnetic regions that allowed coronal holes to develop. We conclude that certain patterns of active nests in Carrington coordinates serve as precursors to coronal holes and fast solar wind streams, providing a pathway to improve space weather predictions.

\section{Data products}\label{sect:2}

\subsection{Carrington maps}\label{sect:2:maps}

We used data from the Solar Dynamics Observatory \citep[SDO;][]{pesnell2012solar} up to August 2026. Synoptic Carrington maps of the radial photospheric magnetic field were taken from the Helioseismic and Magnetic Imager \citep[HMI;][]{scherrer2012helioseismic} with the polar field correction applied \citep{sun2018polar}. Each magnetogram (\textit{hmi.synoptic\_mr\_polfil\_720s}) contains multiple full-disk images spanning one Carrington rotation (CR). We extracted the unsigned magnetic flux within the strong- (> 50 G) and weak-field (< 50 G) regions, used in Sect. \ref{sect:3}. The magnetograms were decomposed into spherical harmonics, following the methodology from \citet{finley2023evolution}, in order to assess the magnetic field strength at different spatial scales and to reconstruct the coronal magnetic field using the potential field source surface (PFSS) model with a source surface radius of 2.5~$R_{\odot}$ \citep{altschuler1969magnetic, schrijver2003photospheric}. The positive and negative poles of the dipole field were obtained from the spherical harmonic decomposition \citep[see also][]{finley2024nested}, with the line joining these points defining the dipole axis.

\begin{figure*}[h!]
    \centering
    \includegraphics[trim=0cm 0cm 0cm 0cm, clip, width=0.9\textwidth]{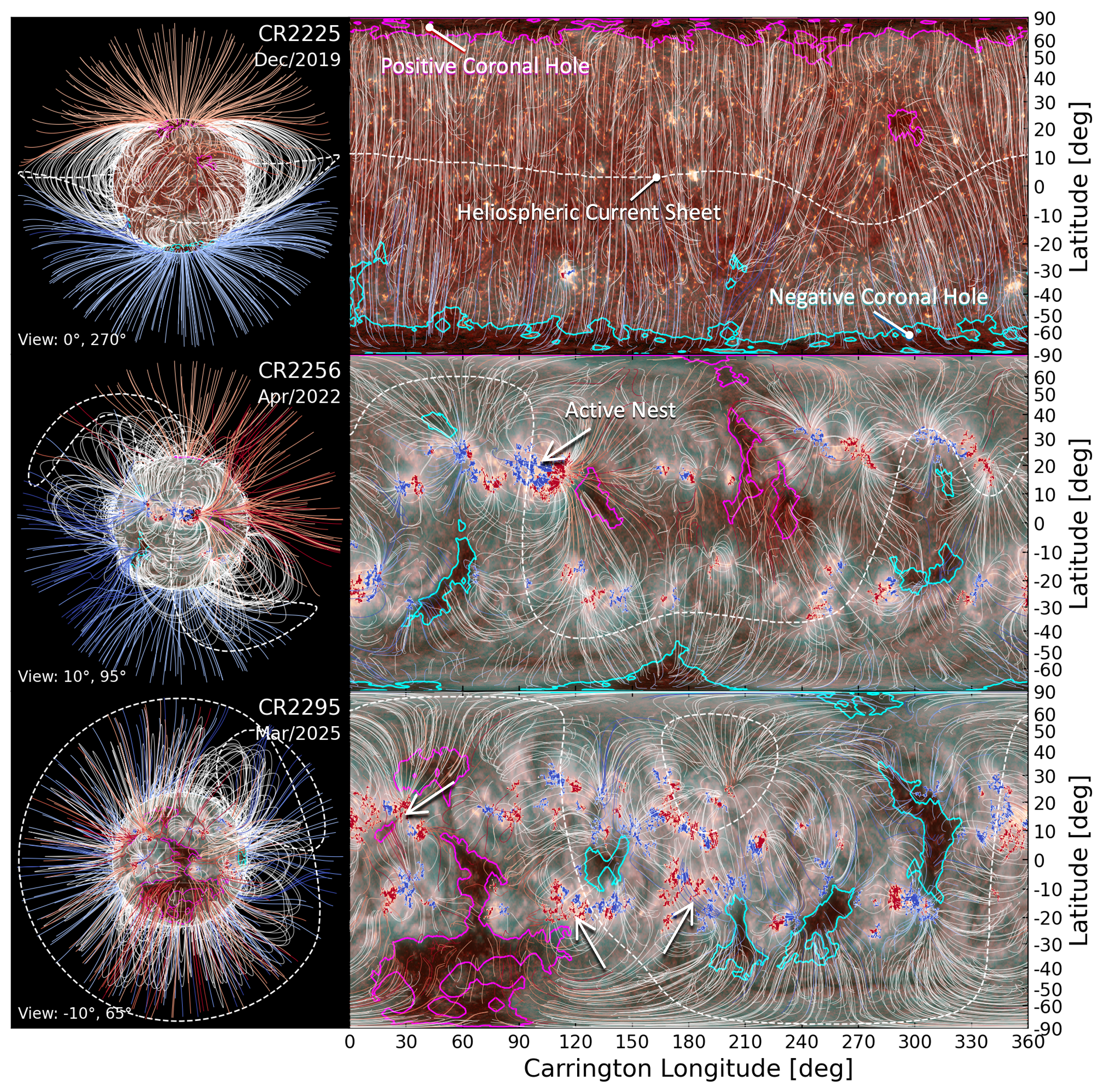}
    \caption{Composite SDO/AIA maps and PFSS extrapolations for three selected CRs in solar cycle 25. \textit{Left:} 3D renderings with the viewing angle annotated. \textit{Right:} Data in Carrington coordinates. Magnetic field elements stronger than 50~G in the SDO/HMI magnetograms are shown in red (positive) or blue (negative). Magnetic field lines from the PFSS are coloured red (open positive), blue (open negative), or white (closed field lines). Dashed white lines highlight the HCS. Solid contours outline coronal holes in magenta (positive) and cyan (negative), identified using the methodology from Sect.~\ref{sect:2:holes}. }
    \label{fig:exampleCarringtonMaps}
\end{figure*}

\begin{figure}[htbp]
    \centering
    \includegraphics[trim=0cm 0cm 0cm 0cm, clip, width=0.5\textwidth]{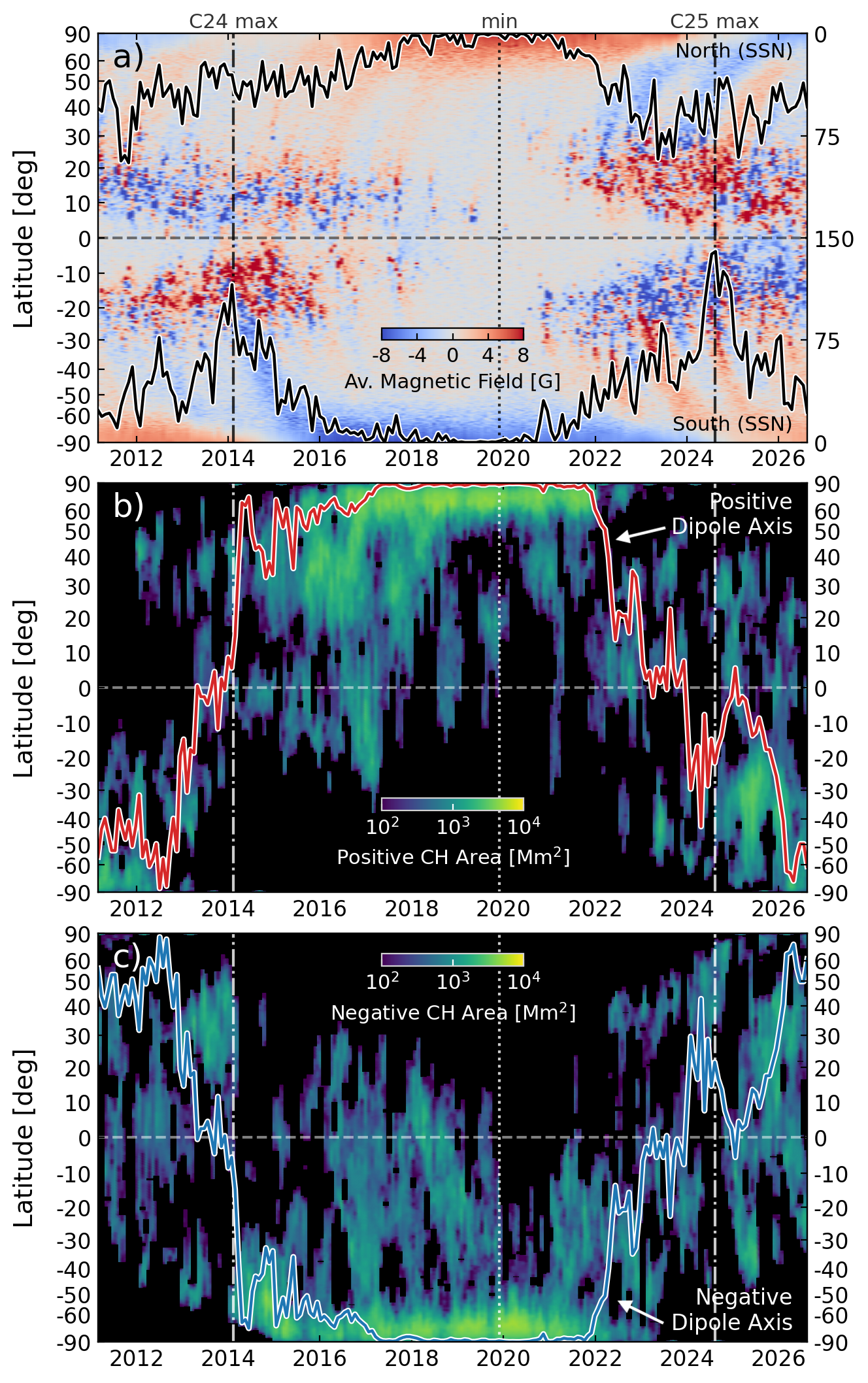}
    \caption{Longitudinally averaged magnetograms and total coronal hole
    surface areas derived from SDO/HMI and SDO/AIA data. \textit{Panel a:} Magnetic butterfly-diagram with the monthly hemispheric sunspot numbers taken from SILSO plotted over each hemisphere. The northern sunspot number y-axis is inverted. \textit{Panels b and c:} Latitudinal distribution of positive and negative polarity coronal hole areas, with the corresponding dipole axes shown in red and blue, respectively. Dot-dashed vertical lines mark the sunspot maxima of solar cycles 24 and 25, while a dotted vertical line denotes the solar minimum (December 2019).}
    \label{fig:coronalHoleButterfly}
\end{figure}

Synoptic Carrington maps of EUV emission were produced from full-disk images taken by the Atmospheric Imaging Assembly \citep[AIA;][]{lemen2012atmospheric}. Specifically, the 304\AA, 211\AA, and 193\AA\ wavelength channels were chosen to increase the contrast of coronal holes with the quiet-Sun and filaments. The SunPy package \citep{sunpy_community2020} was used to re-project and combine daily full-disk images into the Carrington coordinate frame, using a Gaussian weight kernel that favoured observations closer to the sub-solar point. The data were normalised and coloured using standard SDO/AIA scales. The three channels were then combined into a single composite map using a screen-blending technique, with the 193\AA\ and 211\AA\ channels weighted at 50\% relative to the 304\AA\ channel to prevent saturation and balance the coronal emission. 

Figure \ref{fig:exampleCarringtonMaps} contains examples of the composite SDO/AIA maps and PFSS extrapolations of the SDO/HMI magnetograms. At solar minimum (top panel), the large-scale magnetic field was dipolar and dark coronal holes were located at the rotational poles. During the rising phase (middle panel), the emergence of an active nest in 2022 inclined the dipole axis and warped the HCS \citep[for details, see][]{finley2025prolific}. At this time, both bright active regions and dark coronal holes were visible at low-latitudes. Just after solar maximum in 2025 (bottom panel), coronal holes covered large areas around the equator and the large-scale magnetic field resembled a tilted-dipole \citep[][]{finley2026active, rivera2026dipole}. The large positive coronal hole at 70$^{\circ}$ Carrington longitude was a persistent feature spanning almost two years, as discussed in Sect. \ref{sect:3:hole}.

\subsection{Coronal hole identification}\label{sect:2:holes}

From the composite SDO/AIA maps, we identified coronal holes and extracted their areas, recording their corresponding magnetic fluxes from the SDO/HMI magnetograms. Coronal holes are routinely identified in SDO/AIA data using intensity thresholds applied to one or multiple wavelength channels \citep[for a comparison of methods, see][]{reiss2024community}. However, this remains challenging because coronal hole boundaries vary between wavelengths, are obscured by bright active regions or large viewing angles, and are easily confused with filament channels. Despite this, long-term studies have successfully characterised the evolution of coronal hole morphologies and areas during previous solar cycles \citep{de2011evolution, hamada2018automated, heinemann2019statistical}. Motivated by the space weather risk posed by fast wind streams \citep[e.g.,][]{milovsic2025case}, methods for coronal hole detection continue to evolve, adopting increasingly sophisticated machine learning techniques \citep{jarolim2021multi, inceoglu2022identification}. 

To extract coronal hole areas from the composite SDO/AIA maps, we first created an exclusion mask using the strong-field regions and polarity inversion lines (PILs) from the SDO/HMI magnetograms. We performed a binary dilation on the exclusion mask to expand these features up to a few heliographic degrees, roughly matching the scale of filaments in the EUV maps. Areas identified in this mask were excluded. Next, we created an open-field mask using the PFSS source regions from the CR-1, CR, and CR+1 extrapolations. We dilated the open-field mask by tens of degrees to account for the known discrepancy between PFSS sources and coronal holes in EUV imagery \citep[e.g.][]{lowder2017coronal}. Each composite SDO/AIA map was then flattened, and an intensity threshold of 0.55 times the map's median value was applied. This threshold, combined with the exclusion mask, provided the initial detection of coronal hole ``cores''. These cores were required to overlap with the open-field mask and be sufficiently dark relative to their surroundings (defined as having at least 20\% of their area darker than the local background). The cores were then geometrically expanded until they reached a secondary threshold (or the exclusion mask). The secondary threshold was adaptive to account for both local intensity variations and heliographic latitude, scaling from 1.15--1.36 times the primary threshold in quiet equatorial regions up to 1.4--2.4 times near the poles and bright active regions. This methodology shares similarities with previous multi-threshold region-growing algorithms \citep[e.g.,][]{caplan2016synchronic}.

Lastly, we required that the dominant magnetic field polarity in each coronal hole matched the open-field mask, and that their signed magnetic flux exceeded 20\% of the unsigned magnetic flux (i.e. the regions contained a distinct flux imbalance). For regions that failed, we iteratively trimmed the brightest EUV pixels to remove potential contamination from active regions, and then re-evaluated the magnetic field criteria. To account for genuine coronal holes that the open-field mask failed to capture, a final step identified isolated, dark regions with significantly unipolar magnetic fields (signed flux over 45\% of the unsigned flux). For each CR, we recorded the surface area of positive and negative coronal holes, as well as the magnetic flux from the SDO/HMI magnetograms. Examples of the coronal hole areas identified using this methodology are shown in Figure \ref{fig:exampleCarringtonMaps}. While more complex detection methods exist in the literature, this approach yielded sufficient accuracy for this study.

Figure~\ref{fig:coronalHoleButterfly} summarises the latitudinal evolution of the photospheric magnetic field (panel a) and total coronal hole area (panels b and c) extracted from SDO data up to August 2026. The SDO/HMI magnetograms were averaged in longitude to produce a magnetic butterfly-diagram. The monthly hemispheric sunspot numbers were taken from the World Data Center SILSO at the Royal Observatory of Belgium\footnote{Data accessed August 2026: \url{https://www.sidc.be/SILSO/datafiles}}. For each CR map, the coronal hole area was summed in longitude. The dipole component of the Sun's magnetic field was imprinted in the coronal hole distribution, as highlighted by the comparison with the dipole axis. However, during the declining phase of cycle 24, many coronal holes were not aligned with the dipole. As the polar fields recovered, the dipole axis naturally moved closer to the rotational axis, but the decay of active regions continued to generate equatorial coronal holes around the magnetic butterfly-pattern until solar minimum. As the HCS flattened and returned to the ecliptic plane, it nonetheless remained warped around strong active regions. Trans-equatorial coronal holes extended up to $30^{\circ}$ into the opposite polarity hemisphere, most notably from 2016 to 2017. Most of the coronal hole area was located in the hemisphere with matching polar field polarity, typically as polar coronal hole extensions. These high-latitude coronal holes will be accessible to Solar Orbiter's inclined orbit \citep{zouganelis2020solar}, providing a pathway to investigate the fast solar wind emerging from the Sun's poles during the declining phase and upcoming solar minimum. 

Unlike cycle 24, the dipole axis in cycle 25 moved back toward the equator after the 2024 solar maximum. This suggests that, with cycle 24 being particularly weak, the emerging flux from cycle 25 was already sufficient to reverse the polar fields by 2023 (see poleward surges of unipolar flux in Fig.~\ref{fig:coronalHoleButterfly}). Surface flux transport models using data up to August 2023 had forecast that the polar field would reverse in mid to late 2024 \citep{jha2024predicting}. Subsequent flux emergence strengthened the non-axisymmetric component of the dipole, pulling the dipole axis back toward the equator. In addition, Appendix~\ref{ap:cycle24} (see Fig.~\ref{fig:averageCHbfield}) shows that the average magnetic field strength in coronal holes increased by $\sim1$~G from cycle 24 to 25. As these unipolar magnetic field regions are transported poleward by meridional circulation, this hints that the polar fields, and by extension the following cycle's amplitude, could be stronger than the last \citep[see also][predicting stronger polar fields than the last minimum]{wang2026recent}.

\begin{figure*}[h!]
    \centering
    \includegraphics[trim=0cm 0cm 0cm 0cm, clip, width=\textwidth]{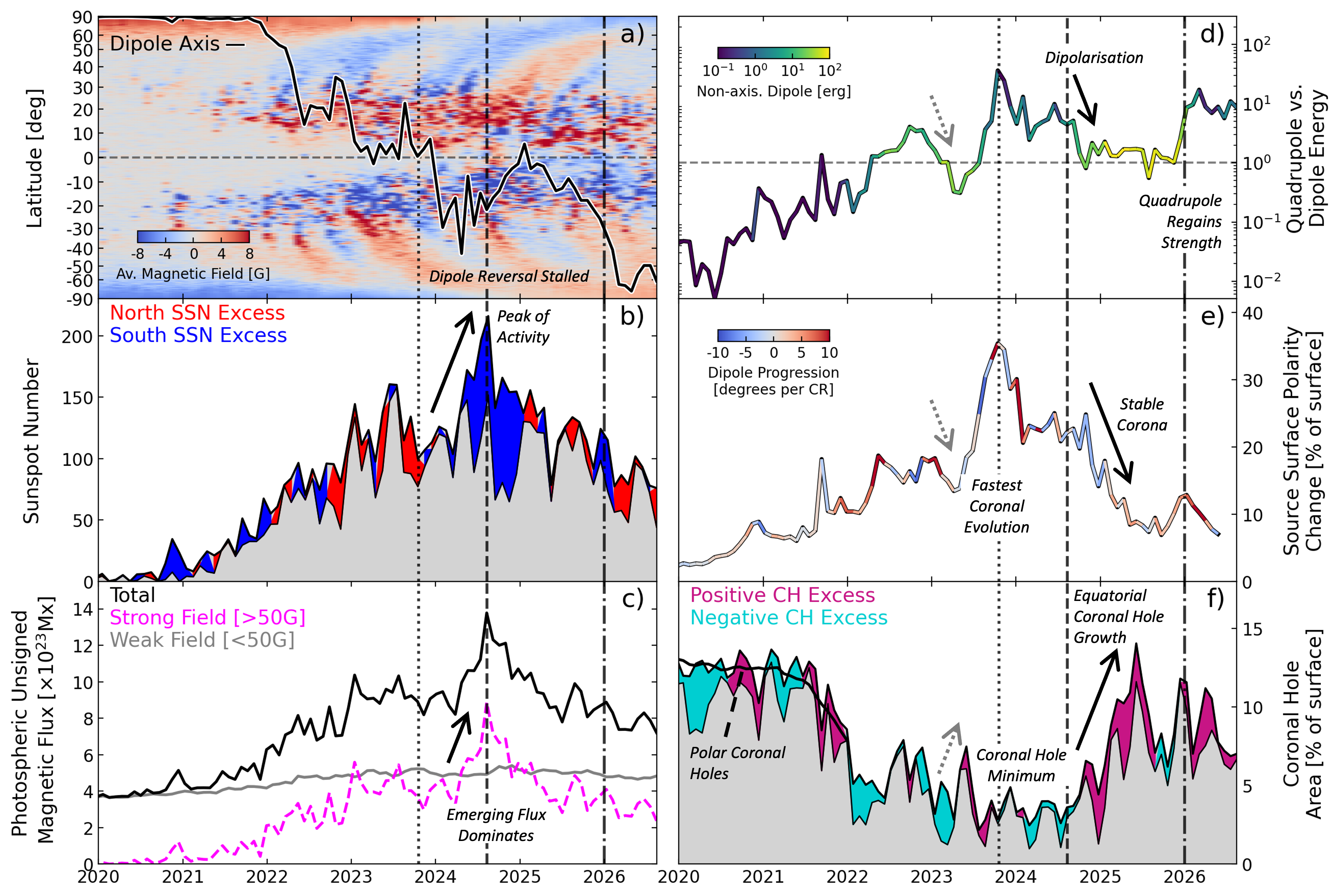}
    \caption{Evolution of the Sun's magnetic field and activity across solar cycle 25 up to August 2026. \textit{Panel a:} Magnetic butterfly-diagram with the dipole's positive pole overplotted. \textit{Panel b:} Monthly sunspot number, highlighting the north-south hemispheric asymmetry. \textit{Panel c:} Total unsigned magnetic flux from SDO/HMI and its partitioning between strong- and weak-field regions. \textit{Panel d:} Quadrupole-to-dipole energy ratio from spherical harmonic decomposition. \textit{Panel e:} Coronal variability, defined as the fraction of the PFSS model source surface area whose magnetic field polarity changed across a sliding 6-map ($\pm 3$ CRs) window. \textit{Panel f:} Total coronal hole area, highlighting imbalances between the positive and negative polarities. To account for the annual viewing-angle bias during solar minimum, the smoothed total area of polar coronal holes is shown with a solid black line from 2020 to 2022. In each panel, vertical dotted, dashed, and dot-dashed lines mark the peak of coronal variability (October 2023), the solar activity maximum and onset of dipolarisation (August 2024), and the transition back to a quadrupolar-dominated state (December 2025), respectively.}
    \label{fig:cycleSummary}
\end{figure*}

\begin{figure*}[htbp]
    \centering
    \includegraphics[trim=0cm 0cm 0cm 0cm, clip, width=\textwidth]{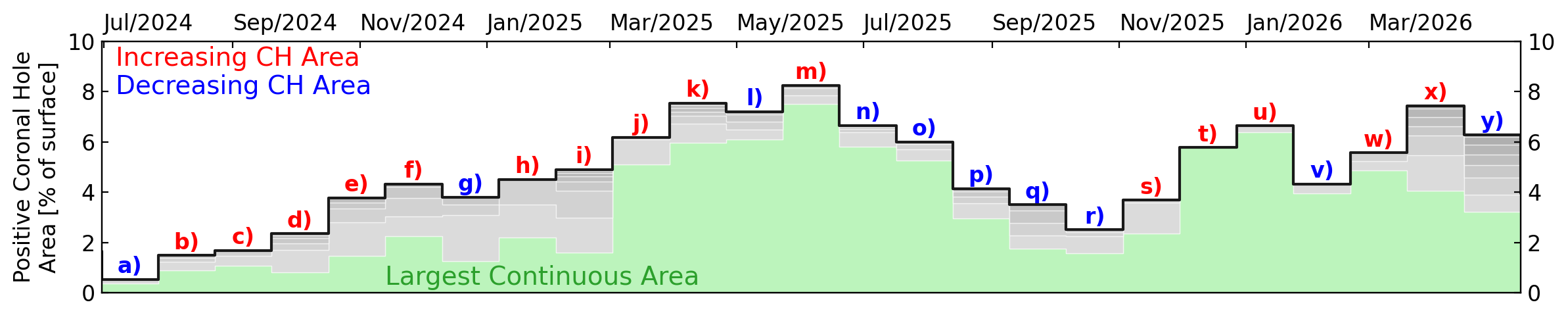}
    \includegraphics[trim=0cm 0cm 0cm 0cm, clip, width=\textwidth]{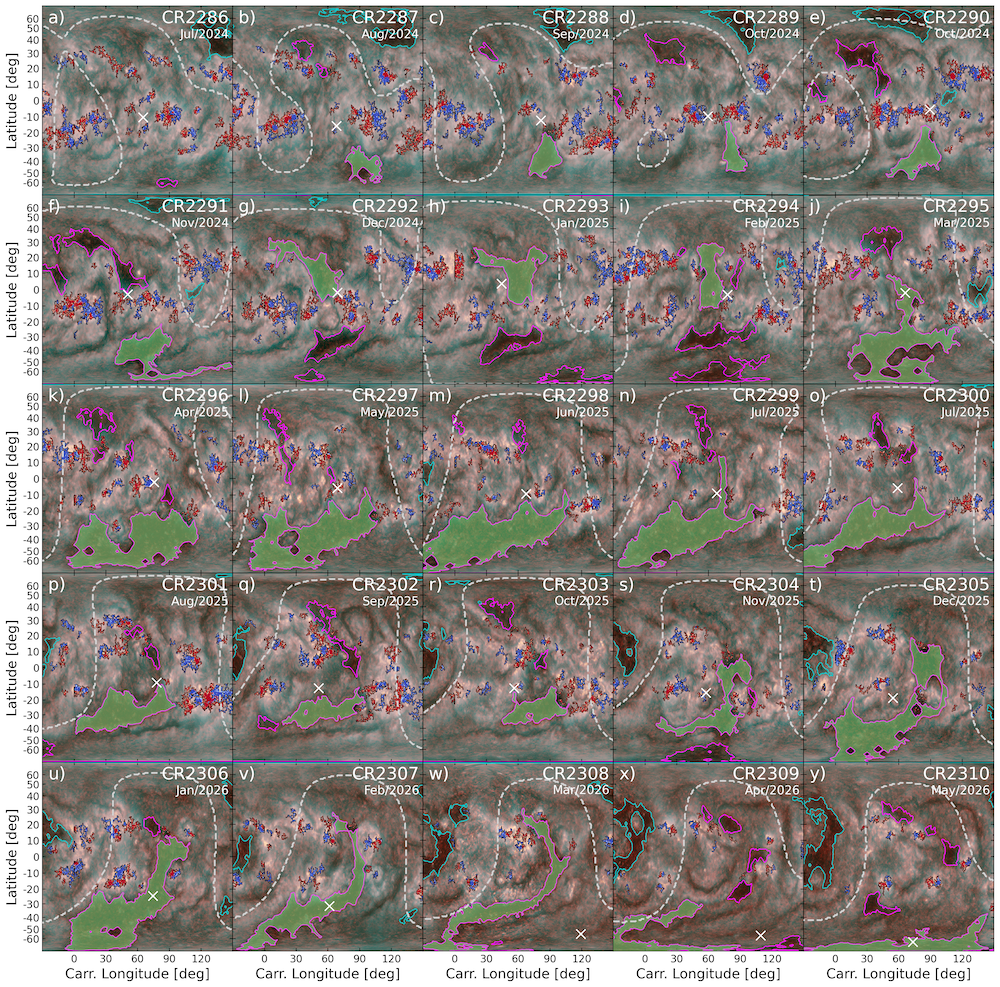}
    \caption{Formation and evolution of a persistent non-axisymmetric coronal hole from July 2024 until May 2026. \textit{Top panel:} Sum of the surface area covered by positive coronal holes, with the largest continuous area highlighted in green. The letter markers (a to y) are coloured red or blue depending on the change in total coronal hole area per CR. \textit{Bottom panel:} Similar to Fig. \ref{fig:exampleCarringtonMaps}, the sub-panels (a) to (y) show the corresponding composite SDO/AIA map (background), strong-field elements (red and blue areas), coronal hole areas (magenta and cyan solid contours), and the HCS location (dashed line). The largest continuous positive coronal hole is filled green and the dipole's positive pole is marked with a white cross.}
    \label{fig:coronalHoleSummary}
\end{figure*}

\begin{figure*}[h!]
    \centering
    \includegraphics[trim=0cm 0cm 0cm 0cm, clip, width=0.9\textwidth]{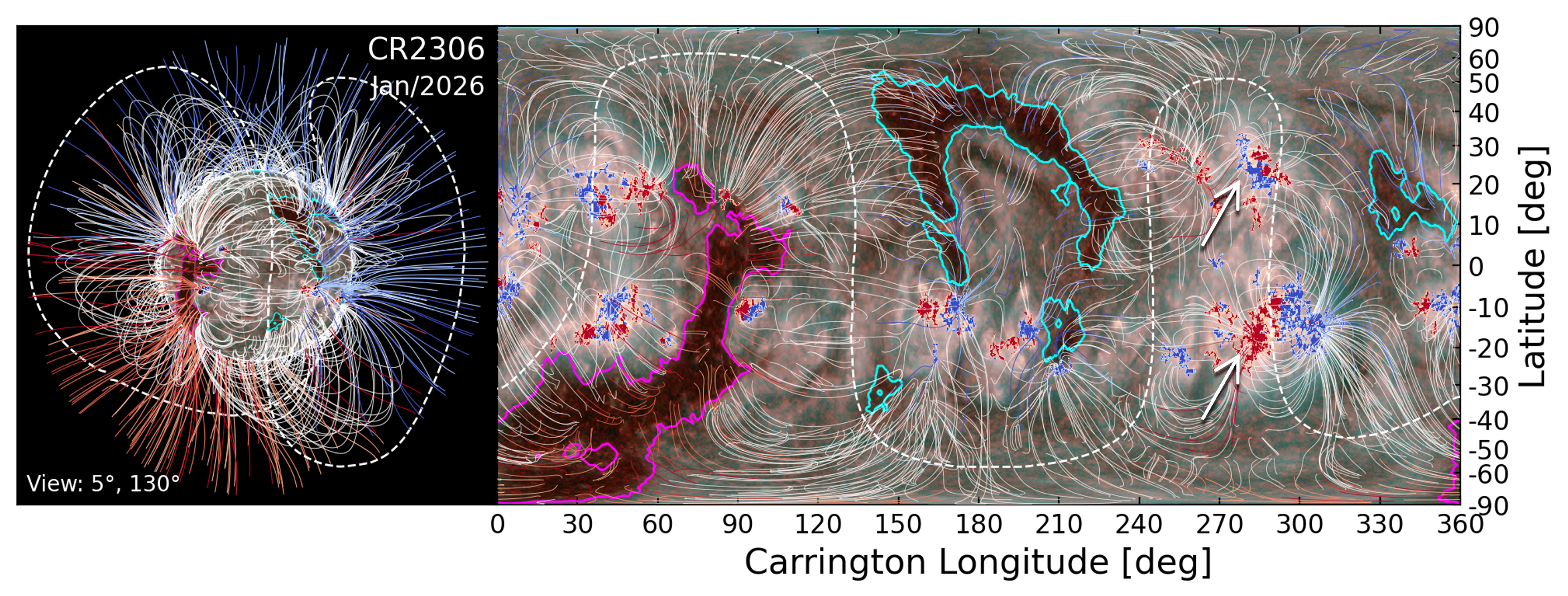}
    \caption{Same as Figure \ref{fig:exampleCarringtonMaps}, but now for CR2306 (January 2026), showing the quadrupole-dominated state. Arrows point towards recently emerged magnetic flux that anchored the HCS around the new positive sector.}
    \label{fig:exampleQuadrupole}
\end{figure*}

\section{Results}\label{sect:3}

\subsection{Overview of solar cycle 25}\label{sect:3:cycle}

Using data from Sect. \ref{sect:2}, Fig. \ref{fig:cycleSummary} outlines the evolution of the Sun's magnetic field and solar activity over solar cycle 25. The magnetic butterfly-diagram and the dipole's positive pole from Fig.~\ref{fig:coronalHoleButterfly} track the emergence and transport of magnetic flux and the coronal magnetic field reversal (panel a). The monthly sunspot number peaked around 210 in August 2024 (panel b) with a maximum unsigned photospheric magnetic flux of $\sim14\times10^{23}$~Mx (panel c). Afterwards, solar activity steadily declined and the active latitudes approached the equator. The ratio of quadrupole-to-dipole energy at the photosphere \citep[see][]{derosa2012solar} grew with increasing solar activity, reaching a peak of $\sim36$ in October 2023 (panel d). This coincided with the largest rotation-to-rotation variability in the coronal magnetic field (panel e), defined as the fraction of the source surface that changed radial magnetic polarity across a sliding six magnetogram ($\pm 3$ CRs) window. Polar coronal holes covered $\sim12\%$ of the Sun's surface at solar minimum, decreasing to a few percent at solar maximum (panel f).

At the recent solar maximum, there was a strong hemispheric asymmetry in the sunspot number (Fig.~\ref{fig:cycleSummary} panel b). \citet{norton2026hemispheric} linked the asymmetry of previous solar cycles to active region nesting. Indeed, \citet{finley2026active} showed that activity in the southern hemisphere increased sharply during 2024 due to recurring flux emergence within three major active nests. The reversal of the dipole field stalled at this time, with the dipole axis moving back towards the Sun's equator as the dominant east-west magnetic field orientation in active nests reinforced the non-axisymmetric dipole. The magnetic flux within strong-field regions peaked alongside the sunspot number, nearly doubling the pre-existing magnetic flux. The quadrupole-to-dipole energy ratio dropped towards unity in the following year (panel d). The ``dipolarisation'' of the coronal field stabilised the PFSS source surface polarity, producing a sector structure that was as steady as the corona in 2021 (panel e). Equatorial coronal holes grew during the tilted-dipole state, covering up to $14\%$ of the solar surface. The total area of positive coronal holes was slightly larger than the negative coronal holes, 8\% versus 6\%, with the positive coronal holes concentrated into one large patch. The negative coronal hole area was divided into a few separate regions (e.g., CR2295 in Fig.~\ref{fig:exampleCarringtonMaps}). This asymmetry was measured in the equatorial solar wind \citep[see][]{rivera2026dipole}, with the negative polarity sector spanning a larger range of Carrington longitudes and split by a pseudostreamer. \citet{zhao2005prediction} found that asymmetry in the positive-negative sector structure was linked to differences in their respective open field regions, such as the mean strength and expansion factor.

The tilted-dipole state ended in December 2025 due to the emergence of active regions in both hemispheres near the Carrington longitude of the dipole's negative pole; further discussed in Sect. \ref{sect:3:quad}. This strengthened the quadrupolar mode, raising the quadrupole-to-dipole ratio back to values last seen in September 2024. Despite this shift, coronal variability remained low in 2026 with coronal holes covering $\sim6\%$ of the solar surface. Examining cycle 25 in more detail, we identified another period  in May 2023 (annotated in Fig.~\ref{fig:cycleSummary}) when the quadrupole-to-dipole ratio decreased sharply. While much shorter, this period was also associated with a decrease in coronal variability and growth in coronal hole area. The link with emerging magnetic flux was less evident for this period, however this was the first time in cycle 25 that the strong-field flux reached a similar strength to the weak-field. Appendix \ref{ap:cycle24} explores a similar sequence of events in cycle 24.

\subsection{Persistent positive coronal hole in the tilted-dipole state}\label{sect:3:hole}

There was a sharp rise in coronal hole area following the maximum of solar cycle 25, with positive coronal holes covering a larger and more continuous area than the somewhat fragmented negative coronal holes; as discussed in Appendix \ref{ap:neghole}. Figure~\ref{fig:coronalHoleSummary} tracks the evolution of non-axisymmetric coronal holes around the tilted-dipole's positive pole ($\sim70^{\circ}$ Carrington longitude) from July 2024 to May 2026, and isolates the largest continuous positive coronal hole area. Initially, small positive coronal holes formed at high latitudes ($>30^{\circ}$) in both hemispheres, connected to decaying active regions (panels a to e). Following the hemispheric east-west orientation for emerging magnetic field \citep{hale1919magnetic}, active regions at 0/360$^{\circ}$ Carrington longitude in the north and 120$^{\circ}$ longitude in the south reinforced the pre-existing non-axisymmetric dipole field and created positive unipolar magnetic regions aligned with the dipole axis. 

Active regions in the southern hemisphere prevented the positive coronal holes in each hemisphere from merging together until March 2025 (panel j). In the north, positive magnetic flux accumulated near the equator, where differential rotation was weak, several degrees per CR, preserving a stable channel for the escaping fast solar wind (panels f to i). Due to the hemispheric north-south orientation of emerging active regions (Joy's law), positive magnetic flux in the southern hemisphere decayed at higher latitudes and experienced more appreciable differential rotation, stretching out the southern coronal hole \citep[as described in][]{wang1990magnetic}. Flux emergence in the southern hemisphere sustained the coronal hole by repeatedly injecting unipolar magnetic regions (panels i to r). Above $\sim30^{\circ}$ latitude, surface differential rotation swept both positive and negative unipolar magnetic regions toward the east, with drift rates on the order of $-70^{\circ}$ longitude per CR \citep[consistent with surface rotation, e.g.,][]{snodgrass1983magnetic, komm1993rotation}. Negative unipolar magnetic regions closed off the southernmost extent of the coronal hole by forming filament channels (panels g-i, o-s, and w-x), whereas positive unipolar regions periodically expanded the total coronal hole area \citep[panels j and t; also observed in previous cycles, see][]{navarro1994observational, insley1995differential, harvey2002polar}. The maximum positive coronal hole area was reached in June 2025 (panel m), with the largest continuous region spanning $\sim8\%$ of the solar surface with an average field strength of 4.6~G (total signed flux of $2.3\times10^{22}$~Mx).

Between June and October 2025 (panels m to r), the total coronal hole area steadily decreased despite support from the decay of neighbouring active regions. As the HCS remained constrained by active nests in the north and south, as discussed in Sect. \ref{sect:4}, high-latitude positive unipolar magnetic regions provided additional area for the coronal magnetic field to open into the solar wind. From November 2025 to January 2026 (panel s to u), coronal hole area increased due to the alignment of a positive unipolar magnetic region carried by differential rotation near the south pole. This also coincided with the opening of the positive quadrupole sector over the dipole's negative pole (see Appendix~\ref{ap:neghole}). The high-latitude unipolar magnetic regions were subsequently sheared by differential rotation and partially closed by flux cancellation and filament formation (panels x and y). Positive coronal holes were present at this longitude until at least August 2026, spanning two years (nearly $20\%$ of the 11-year solar cycle). Although its lifetime was significantly longer, the growth and decay behaviours of this coronal hole shares similarities with large coronal holes identified during solar cycle 24 \citep[][]{heinemann2020statistical}.

The fast solar wind from this long-lived coronal hole was measured as close as ten solar radii by the Parker Solar Probe in March 2025 \citep{rivera2026dipole}, directly probing the acceleration region before the stream was measured again at Solar Orbiter and then at 1~au \citep{badman2026direct}. The persistence of this feature made it a recurring target for high-resolution imaging and spectroscopy across multiple Solar Orbiter solar wind connection science campaigns in March 2025 (CR2295), February 2026 (CR2308), and August 2026 (CR2314); data that is currently under investigation.

\begin{figure*}[h!]
    \centering
    \includegraphics[trim=0cm 0cm 0cm 0cm, clip, width=\textwidth]{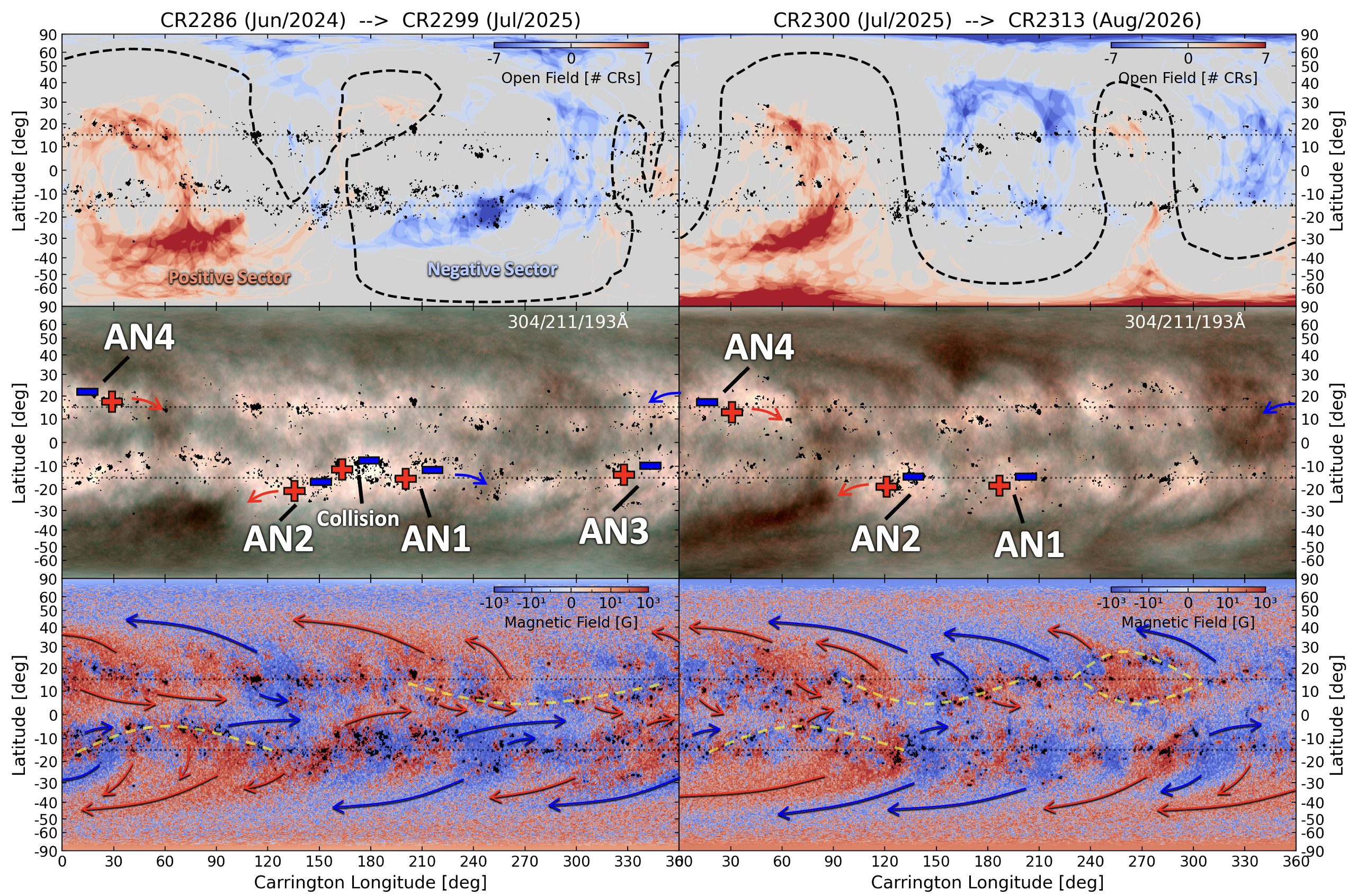}
    \caption{Summary of the Sun's magnetic field and EUV emission in the tilted-dipole state (left) and the quadrupole-dominated state (right). \textit{Top panels:} Distribution of open-field regions from PFSS models using SDO/HMI magnetograms from each interval. Field strengths greater than 1~kG are shown with black filled areas and dotted horizontal lines indicate the mean active latitudes (in all panels). The averaged HCS for each interval is marked with a dashed black line and is visibly anchored above strong-field regions. \textit{Middle panels:} Averaged composite SDO/AIA maps identify persistent activity (bright) and coronal holes (dark). Active nests from \citet{finley2026active} are annotated with their local magnetic field orientation. \textit{Bottom panels:} Sum of SDO/HMI magnetograms in each interval with arrows indicating the prevailing surface transport of unipolar magnetic flux. Dashed yellow curves highlight where the active latitudes were notably warped.}
    \label{fig:AveragedMaps}
\end{figure*}

\subsection{Return to a quadrupole-dominated corona}\label{sect:3:quad}

The quadrupole-to-dipole energy ratio returned to the quadrupole-dominated regime in December 2025. However, the flux emergence that led to this shift began as early as May 2025; visible in Appendix~\ref{ap:neghole}. Figure~\ref{fig:exampleQuadrupole} shows how the coronal magnetic field was organised in January 2026 (CR2306), after being reconfigured by flux emergence across both hemispheres near 270$^{\circ}$ Carrington longitude. The positive coronal hole discussed in Sect.~\ref{sect:3:hole} was largely unaffected by the evolving coronal topology, reaching the second largest area coverage in Fig.~\ref{fig:coronalHoleSummary} (panel u). Whereas the negative coronal hole area had been split into two regions, as the new flux emergence pattern coincided with the longitude of the dipole's negative pole. This explains the asymmetry in coronal hole formation, as the negative axis was repeatedly disrupted by flux emergence events throughout 2025. Both positive and negative coronal hole areas grew temporarily as the corona reconfigured in December 2025. Throughout 2026, the corona remained quadrupole-dominated and, due to declining solar activity, was relatively stable (see Fig.~\ref{fig:cycleSummary}). For this reason, the large-scale magnetic field from previous CRs provided a reasonable predictor of coronal structure observed during the total solar eclipse in August 2026.

\section{Discussion}\label{sect:4}

\begin{figure*}[h!]
    \centering
    \includegraphics[trim=0cm 0cm 0cm 0cm, clip, width=\textwidth]{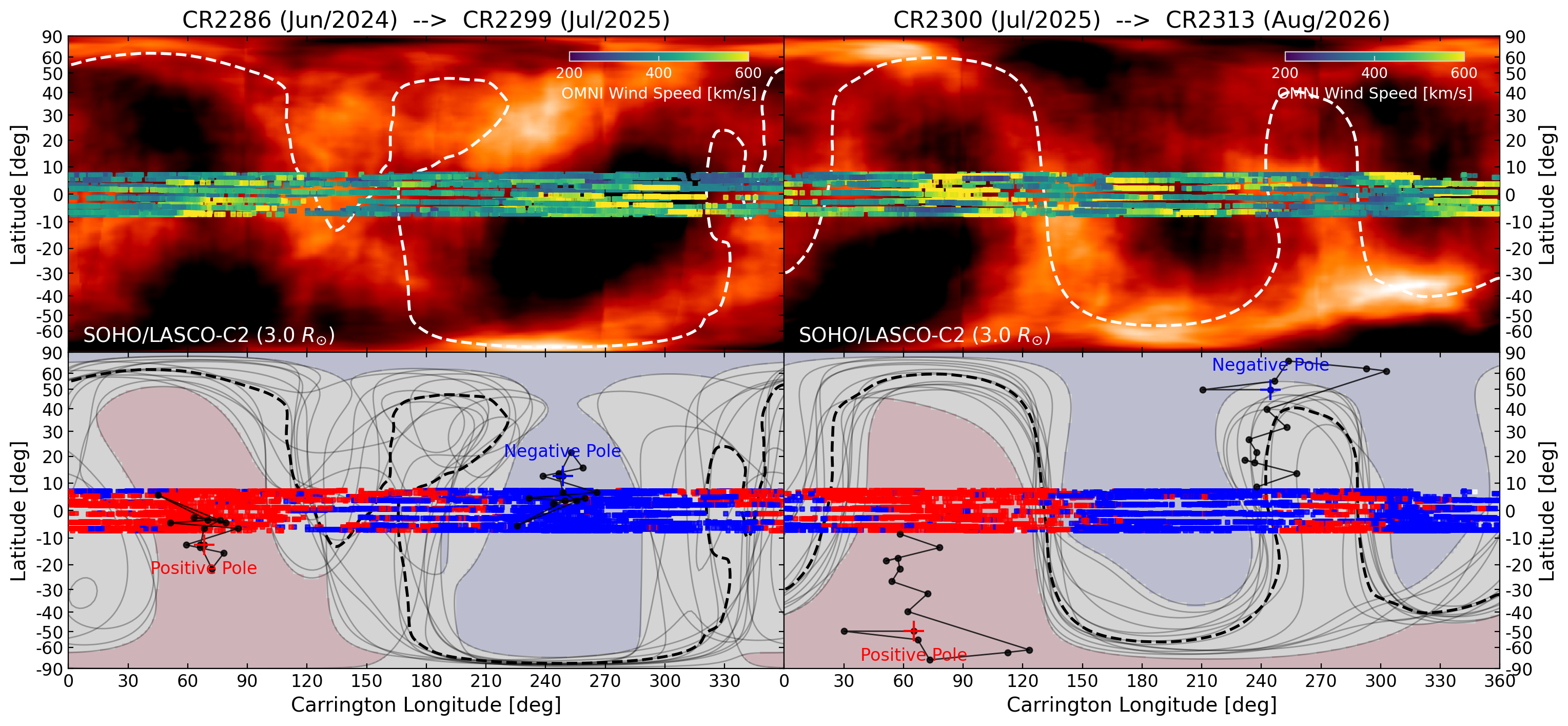}
    \caption{Summary of the solar corona and solar wind at 1~au in the tilted-dipole state (left) and the quadrupole-dominated state (right). \textit{Top panels:} Averaged SOHO/LASCO-C2 scattered white light at three solar radii, overlaid with ballistically back-mapped hourly in situ solar wind speed from the OMNI database. \textit{Bottom panels:} Source surface polarity from PFSS models using SDO/HMI magnetograms spanning each interval. Solid black lines outline the HCS from each CR. Light red and blue areas highlight regions with consistent magnetic field polarity across all CRs in the interval. The hourly in situ radial magnetic field polarity from OMNI and the evolution of the dipole axis are overlaid for each interval. Thick dashed lines show the averaged HCS locations.}
    \label{fig:solarWind}
\end{figure*}

While strong-field regions, such as active nests, are compact and localised, they have a significant affect the global coronal magnetic field by warping the HCS around structures in the low corona \citep{petrie2013low, finley2024nested}. To connect the spatial distribution of strong-field regions with the open-field area, coronal holes, and the global coronal topology, Figure~\ref{fig:AveragedMaps} shows data from two years, June 2024 to July 2025 and then July 2025 to August 2026. In the first year, the large-scale magnetic field resembled a tilted-dipole. In the second year, the corona transitioned into a quadrupole-dominated state. For each configuration, the top panels of Figure~\ref{fig:AveragedMaps} highlight magnetic fields greater than 1~kG and the open-field regions from PFSS extrapolations of the 14 SDO/HMI magnetograms in each interval, highlighting how often open-field was predicted. We also computed the mean source surface polarity to identify the average HCS position (see also Fig.~\ref{fig:solarWind}). We recovered a dipolar and then quadrupolar HCS, shaped by the underlying distribution of strong-field regions. In 2024, three southern hemisphere active nests were identified in \citet{finley2026active}, designated AN1, AN2, and AN3. Two nests (AN1 and AN2) converged over several months and collided in October 2024, leading to increased solar activity. In 2025, two nests re-emerged from the collision with similar characteristics, and so the labels were retained. At this time, a new active nest (AN4) formed in the northern hemisphere, while AN3 dissipated. These active nests are annotated in the middle panels of Fig.~\ref{fig:AveragedMaps}.

The averaged SDO/AIA composite maps from each year are shown in the middle panels of Fig.~\ref{fig:AveragedMaps}, revealing bright areas with persistent activity and dark regions associated with long-lived coronal holes and filament channels. The darkest regions were organised by the large-scale magnetic field and transport of unipolar magnetic flux from decaying active regions. In the tilted-dipole state, flux emergence within AN1, AN2, and their collision product anchored the HCS near to $180^{\circ}$ Carrington longitude, with unipolar magnetic regions from the combined leading and trailing edges contributing to open-field areas in the PFSS models. As the polar fields began to recover in the following year, more open-field appeared at the rotational poles. AN2 continued to anchor the HCS and supplied unipolar magnetic flux to neighbouring open-field regions. In the quadrupole-dominated state, the distribution of open-field in the large positive coronal hole was often split from the southern polar field by negative unipolar magnetic regions forming at the leading edge of AN2. Positive open-field areas formed at $270^{\circ}$ Carrington longitude once the quadrupolar state was established. Similar to the dipolar coronal hole from Sect.~\ref{sect:3:hole}, the quadrupolar open-field was enhanced when unipolar magnetic regions, transported by differential rotation near the poles, aligned with the pre-existing HCS sector structure. This effect was also evident during the rising phase of cycle 25, when the polar coronal holes weakened and the HCS became initially warped \citep[as noted in][for cycle 23]{sanderson2003observations}. 

The bottom row of Fig.~\ref{fig:AveragedMaps} combines the SDO/HMI magnetograms from each year to follow the emerged magnetic flux and its transport via turbulent diffusion, meridional circulation, and differential rotation \citep[for a review, see][]{yeates2023surface}. From this, we identified the dominant pathways for decaying unipolar magnetic flux in each configuration. In some cases, like the persistent coronal hole from Sect.~\ref{sect:3:hole}, decaying unipolar magnetic flux was transported over large distances before cancelling with oppositely oriented field or reinforcing the polar fields. The longitudinal separation of active regions, and especially nests, with their nearest neighbours in the opposite hemisphere affected the formation of trans-equatorial coronal holes \citep[see also][]{heinemann2018three2,heinemann2018three}, while unipolar magnetic regions forming at the leading edge of active regions disrupted the connection between the trailing coronal hole polarity and the recently reversed polar field. The long-term PILs revealed in the bottom panels of Fig.~\ref{fig:AveragedMaps} indicated where polar filaments were likely to form and matched the dark filament structures observed in the averaged EUV emission (middle panels). 

The mean active latitude of the butterfly-pattern in Fig.~\ref{fig:cycleSummary} was around $15^{\circ}$ during this period; however, several active regions departed from the activity belts (annotated in Fig.~\ref{fig:AveragedMaps}). In the tilted-dipole state, southern active regions emerging around the dipole's positive pole were up to $10^{\circ}$ closer to the equator. Similar tipping angles where recovered from magnetograms of previous solar cycles \citep{norton2005recovering}, with numerical models attributing this to instabilities within the toroidal magnetic field \citep{dikpati2021deciphering, raphaldini2023information, dikpati2025mother}. It is unclear if the warping of the active latitudes further accelerated coronal hole growth by affecting the orientation of the emerging field or not. In the quadrupolar state, this trend continued with a warped northern activity belt surrounding the negative coronal hole at $180^{\circ}$ Carrington longitude. In addition, the northern active latitude was split into two branches by flux emergence supporting the quadrupolar sector structure; annotated in Fig.~\ref{fig:exampleQuadrupole}. 

Persistent non-axisymmetric coronal holes in both the tilted-dipole and quadrupole-dominated states had a measurable impact on the solar wind speed distribution. The top row of Figure \ref{fig:solarWind} shows yearly averaged Carrington maps of scattered white-light at three solar radii from the Large Angle and Spectrometric Coronagraph \citep[LASCO;][]{brueckner1995large} on board the Solar and Heliospheric Observatory \citep[SOHO;][]{domingo1995soho} with the corresponding hourly averaged in situ solar wind speed at 1~au from the OMNI database \citep{king2005solar} ballistically back-mapped to the same radial distance. Time periods associated with high mass or magnetic fluxes (above $3\times10^{12}$~g/s and $2\times10^{23}$~Mx) were removed to reduce the potential contamination from coronal mass ejections and co-rotating interaction regions\footnote{This process resulted in $\sim325.8$ days and $\sim338.5$ days of data for the tilted-dipole and quadrupole-dominated states, respectively. More data was removed during 2024 as eruptive activity was higher at solar maximum.}. Although this approach suited our current analysis, future studies could use sector boundary catalogues to more precisely isolate the ambient solar wind \citep[e.g.,][]{heinemann2026detection}. The bottom row of Figure~\ref{fig:solarWind} shows the radial magnetic field polarity in the solar wind at 1~au with the PFSS polarity sector structure for both years. Structure in the in situ data was well represented by the averaged HCS location in both configurations. The dipole axis was relatively stable in the tilted-dipole state, but quickly migrated to the poles upon entering the quadrupole-dominated state.

The longitudinal organisation imposed by the configuration of the Sun's large-scale magnetic field creates periodicities in the solar wind arriving at 1~au \citep{mursula199613}. In the averaged SOHO/LASCO-C2 maps, darker areas of lower electron density correlated with the back-mapped fast solar wind streams. Conversely, brighter regions associated with streamers aligned with the slow solar wind and the average HCS location. In the quadrupolar state, the solar wind emerging from the smaller positive sector at 270$^{\circ}$ longitude had significantly slower speeds than the wind from the larger persistent coronal holes discussed in Sect.~\ref{sect:3:hole}. However, we found that the overall distribution of wind speeds measured at 1~au was shifted towards higher values in the quadrupole-dominated state, with a clear increase between 500 and 700~km/s. The average wind speeds for the tilted-dipole and quadrupole-dominated state were 443~km/s and 481~km/s, respectively, with the latter experiencing approximately 55 more days of wind speeds exceeding 500~km/s at 1~au. The speed distributions were inherently biased due to sampling being restricted to the ecliptic plane. As the tilted-dipole was inclined by five to ten degrees from the solar equator, the fastest wind streams may have been poorly sampled or missed entirely. Nonetheless, in the quadrupole-dominated state the negative polarity coronal holes were more appreciable in the averaged EUV emission (see Fig.~\ref{fig:AveragedMaps}), which could have contributed to the development of the fast solar wind. The asymmetry between positive and negative coronal hole area also reduced from $4\%$ of the solar surface down to $\sim1\%$ (see Fig.~\ref{fig:cycleSummary}).

Shifts between dipolar and quadrupolar coronal morphologies have been identified in previous solar cycles. During the first activity peak of cycle 23, there was a brief period of dipolarisation as the dipole reversed around May 2000, returning to a quadrupole-dominated state in early 2001 \citep{sanderson2003observations}. At the second activity peak in 2003, active region nesting enforced a tilted-dipole state again, returning to a warped quadrupolar HCS in 2004. Cycle 24 was weaker than cycle 25 (see Appendix~\ref{ap:cycle24}), but also experienced nesting at solar maximum in the southern hemisphere that rejuvenated the dipole component of the Sun's large-scale field \citep{sheeley2015recent} and triggered the formation of coronal holes \citep{yoshida2026temporal}, increasing the amount of open magnetic flux in the heliosphere \citep{heinemann2024origin}.

\section{Conclusions}\label{sect:conc}

This study connected the strength of flux emergence within active nests to morphological changes in the Sun's large-scale magnetic field and the formation of persistent non-axisymmetric coronal holes. We combined composite maps of EUV emission from SDO/AIA  with magnetograms from SDO/HMI to track the evolution of coronal holes and the large-scale magnetic field across solar cycle 25. Recurring flux emergence at solar maximum overwhelmed the pre-existing background field and triggered the dipolarisation of the coronal magnetic field and solar wind outflow, as measured in situ by the current suite of heliospheric missions \citep{rivera2026dipole}. In this configuration, active nests stabilised the large-scale magnetic field and supported the formation of equatorial coronal holes. These non-axisymmetric coronal holes were similar in size to the polar coronal holes at solar minimum, covering up to $14\%$ of the solar surface. The fast solar wind emerging from these regions was sampled by Parker Solar Probe as close as ten solar radii  \citep{badman2026direct}, probing the nature of its acceleration. 

Coronal holes were observed in EUV near the tilted-dipole's positive pole ($\sim$70$^{\circ}$ Carrington longitude) for almost two years; nearly $20\%$ of the 11-year solar cycle. The yearly averaged SOHO/LASCO-C2 coronagraph data and solar wind speed at 1~au from the OMNI database closely reflected the dominant coronal topology, with low density fast wind streams over the tilted-dipole's positive and negative poles in 2025. This highlighted the relative stability of the global coronal structure over these timescales despite increased activity at solar maximum. Flux emergence in both hemispheres near the Carrington longitude of the dipole's negative pole ($\sim$270$^{\circ}$ Carrington longitude) transformed the tilted-dipole corona into a quadrupole-dominated state in 2026 (ahead of the total solar eclipse over the North Atlantic and western Europe in August). While the persistent positive coronal hole was largely unaffected by this topological shift, the fragmented negative coronal holes stabilised at two distinct longitudes, and the average wind speed measured at 1~au increased. Earth experienced 55 more days per year with wind speeds exceeding 500~km/s in the quadrupole-dominated state than in the tilted-dipole state. Future work could use Solar Orbiter's inclined orbit to investigate how sampling restricted to the ecliptic plane biased these wind speed distributions and assess how the coronal magnetic field controls the total solar wind mass-loss and angular momentum-loss rates \citep{finley2025reconstructing, badman2025multispacecraft}.

Fast wind streams and their associated compression regions represent a substantial risk to infrastructure at 1~au \citep{tsurutani2006corotating, richardson2018solar, grandin2019properties}. The spatial distribution of active nests relative to one another, and to the pre-existing large-scale field, combined with their east-west polarity, encodes the likelihood of coronal hole formation and the birth of persistent fast wind streams. Furthermore, coronal variability appears in opposition to the formation of coronal holes, making it a key metric to follow during the cycle. As discussed in \citet{finley2026active}, the distribution of activity across the solar surface is expected to be intrinsically linked to the progressive depletion of the internal toroidal magnetic field. Consequently, not all flux emergence patterns are equally likely, making the Sun's flux emergence history a valuable tool for bridging space weather and space climate.

\begin{acknowledgements}
AJF acknowledges support through the European Space Agency (ESA) Research Fellowship in Space Science.
MT acknowledges the Austrian Science Fund (FWF) [10.55776/PAT2280725 and 10.55776/PAT1560825].
Data supplied courtesy of the SDO/HMI and SDO/AIA consortia. SDO is the first mission to be launched for NASA's Living With a Star (LWS) Program.
Data manipulation was performed using the numpy \citep{2020NumPy-Array}, scipy \citep{2020SciPy-NMeth}, and pySHTOOLS \citep{wieczorek2018shtools} python packages.
Figures in this work are produced using the python package matplotlib \citep{hunter2007matplotlib}. 
\end{acknowledgements}

%
%

\bibliographystyle{aa}
\bibliography{adam}

\begin{appendix}
\nolinenumbers
\onecolumn
\section{Comparison with solar cycle 24}\label{ap:cycle24} 

\begin{figure*}[htbp]
    \centering
    \includegraphics[trim=0cm 0cm 0cm 0cm, clip, width=\textwidth]{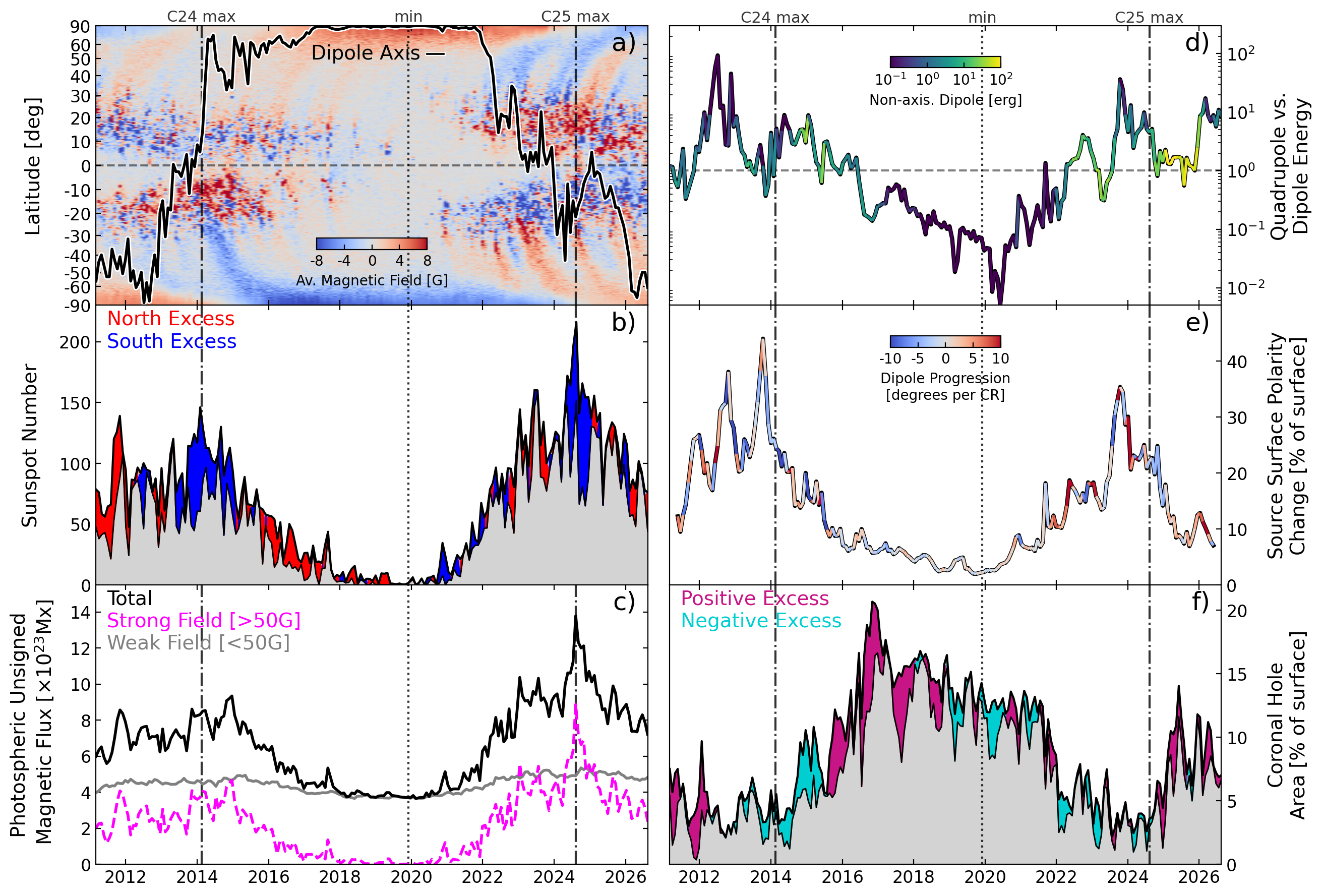}
    \caption{Same as Fig.~\ref{fig:cycleSummary}, but now displaying the full SDO dataset from February 2011 to August 2026. Dot-dashed vertical lines mark the sunspot maxima of solar cycles 24 and 25, while a dotted vertical line denotes the solar minimum (December 2019).}
    \label{fig:SDOera}
\end{figure*}

\begin{figure}[htbp]
    \centering
    \includegraphics[trim=0cm 0cm 0cm 0cm, clip, width=0.9\textwidth]{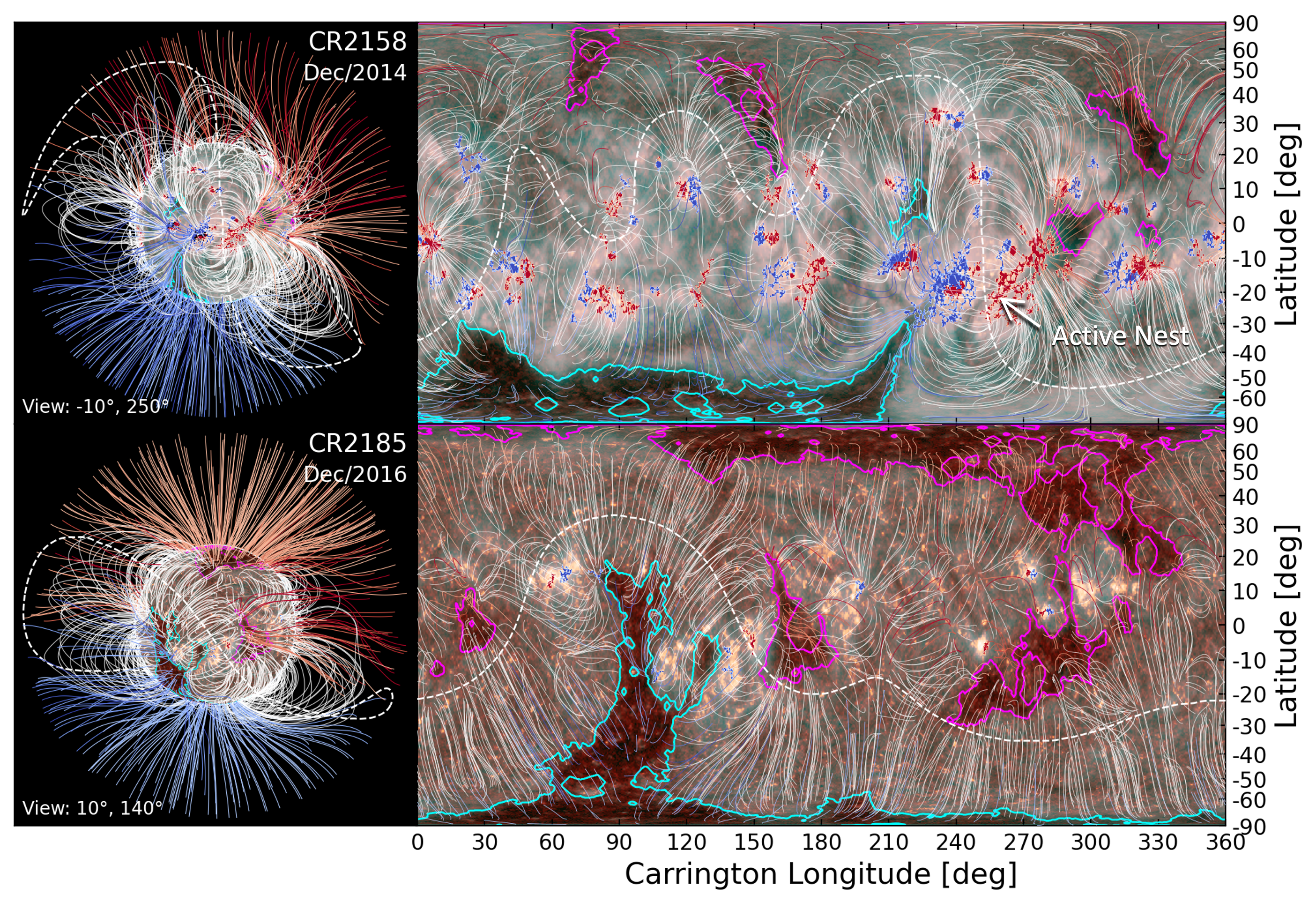}
    \caption{Same as Fig.~\ref{fig:exampleCarringtonMaps}, but now with two selected CRs in solar cycle 24.}
    \label{fig:exampleCycle24}
\end{figure}

Periods of pronounced active region nesting have been identified during previous solar cycles \citep[e.g.,][]{norton2025moderate}, facilitated by the growing volume of ground- and space-based observations. Notably, there was a striking similarity between the solar maxima of cycles 24 and 25. Figure~\ref{fig:SDOera} expands the results from Sect.~\ref{sect:3:cycle} to encompass the entire SDO dataset. Solar activity peaked in the southern hemisphere in 2014, mirroring the behaviour of cycle 25 in 2024 (panels a and b). Nesting in 2014 rejuvenated the dipole component of the Sun's photospheric magnetic field \citep{sheeley2015recent}, strengthening the non-axisymmetric field (panel d). However, unlike cycle 25, the HCS remained highly warped with a strong quadrupole-to-dipole ratio (see CR2158 in Fig.~\ref{fig:exampleCycle24}). This is likely a consequence of the nest's weaker field strength relative to the background field, compounded by the presence of only a single nest in the southern hemisphere. In contrast, cycle 25 had nesting in both hemispheres in 2025 that reinforced the tilted-dipole state. Coronal variability dropped towards the end of 2013, similar to the timing observed in cycle 25 (panel e), with the total coronal hole area increasing shortly after (panel f).

In 2014, nesting supplied negative unipolar flux directly to the newly forming southern polar coronal hole while directing positive flux toward the equator \citep{yoshida2026temporal}. This was accompanied by a sharp rise in the open magnetic flux in the heliosphere, as measured from 1~au \citep{heinemann2024origin}. Coronal hole area began expanding in 2015, growing from a few percent of the solar surface to nearly 10\%, driven primarily by the growth of negative polarity coronal holes. Between 2016 and 2017, this total area reached a maximum of approximately 20\% of the solar surface. The peak area exceeded the polar coronal hole area at solar minimum, which settled closer to 12\%. The surge in coronal hole area following the 2014 solar maximum coincided with the HCS moving back towards the ecliptic and the quadrupole-to-dipole ratio dropping below unity. As the non-axisymmetric dipole component was still strong at this time, the coronal holes were organised into two large patches (see CR2185 in Fig.~\ref{fig:exampleCycle24}). With unipolar magnetic regions present at both the equator and the poles, we observed the shearing and re-alignment behaviour that periodically increased the area of the persistent coronal hole studied in Sect.~\ref{sect:3:hole}. In the following years, the positive polarity coronal hole area exceeded the negative area due to additional flux emergence in the northern hemisphere. This supplied positive unipolar flux to the high-latitude coronal holes, and was correlated with the positive poleward surge during 2016, visible in Fig.~\ref{fig:SDOera} (panel a).

\begin{figure}[htbp]
    \centering
    \includegraphics[trim=0cm 0cm 0cm 0cm, clip, width=0.6\textwidth]{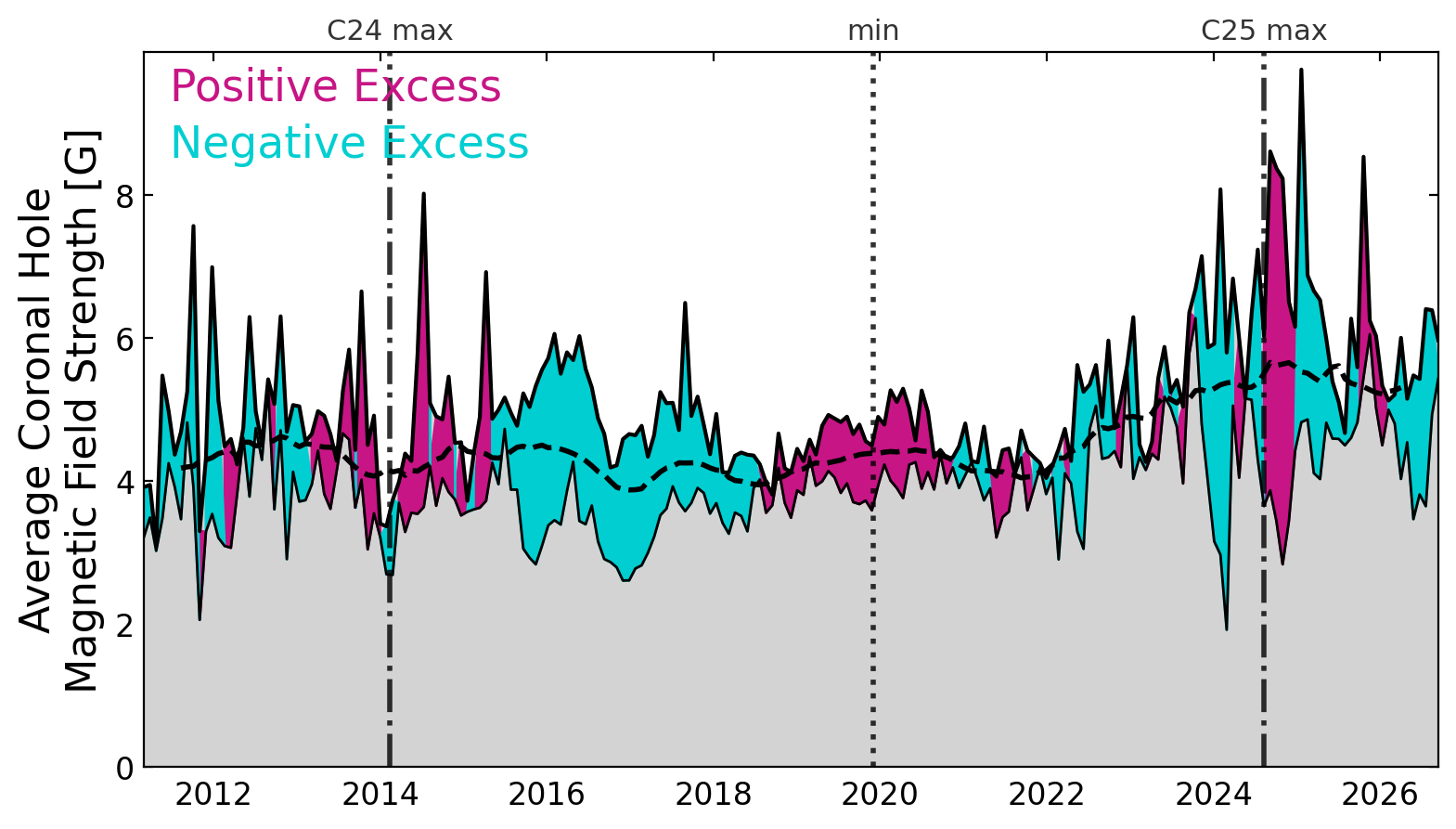}
    \caption{Average magnetic field strength within coronal holes. Colours highlight when one polarity has a stronger average field than the other. Dashed line traces the 6-month rolling average field strength in both positive and negative coronal holes. Vertical lines match Fig.~\ref{fig:SDOera}.}
    \label{fig:averageCHbfield}
\end{figure}

There was a distinct asymmetry between positive and negative coronal holes during the declining phase of cycle 24. From 2016 to 2017, positive coronal holes occupied a larger surface area but exhibited a weaker average field strength (see Fig.~\ref{fig:averageCHbfield}). Specifically, the average field strength in negative coronal holes was stronger from 2015 to 2019, after which the positive polarity dominated until the polar coronal holes closed in 2022. As the polar fields stabilised around solar minimum, the asymmetry in coronal hole area diminished, leaving the smaller annual variations caused by Earth's viewing angle. The systematic differences in average field strength, and by extension expansion factors, during the declining and minimum phases likely caused a north-south asymmetry in the average HCS location \citep[as described in][]{zhao2005prediction}. Comparing the declining phases of cycles 24 and 25, the total non-axisymmetric coronal hole area appears larger in cycle 25, as is the average magnetic field strength (by nearly 1~G). If these unipolar magnetic regions are advected into the polar fields, it could suggest that cycle 26 will be stronger than the previous two, given the well-established correlation between the polar field strength at minimum and the amplitude of the following solar cycle \citep[see][and references therein]{nandy2021progress}.

\newpage
\onecolumn
\section{Fragmented negative polarity coronal holes and the quadrupole emergence}\label{ap:neghole} 

\begin{figure*}[htbp]
    \centering
    \includegraphics[trim=0cm 0cm 0cm 0cm, clip, width=\textwidth]{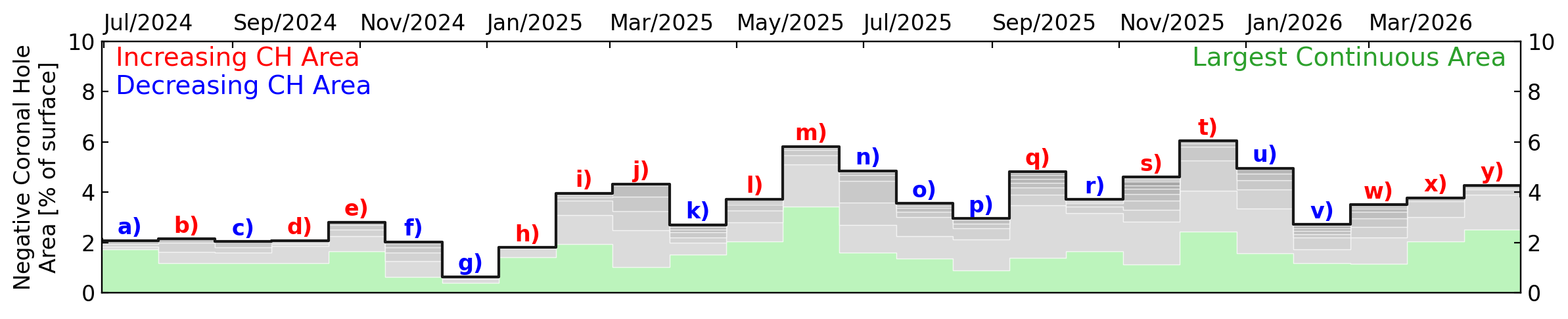}
    \includegraphics[trim=0cm 0cm 0cm 0cm, clip, width=\textwidth]{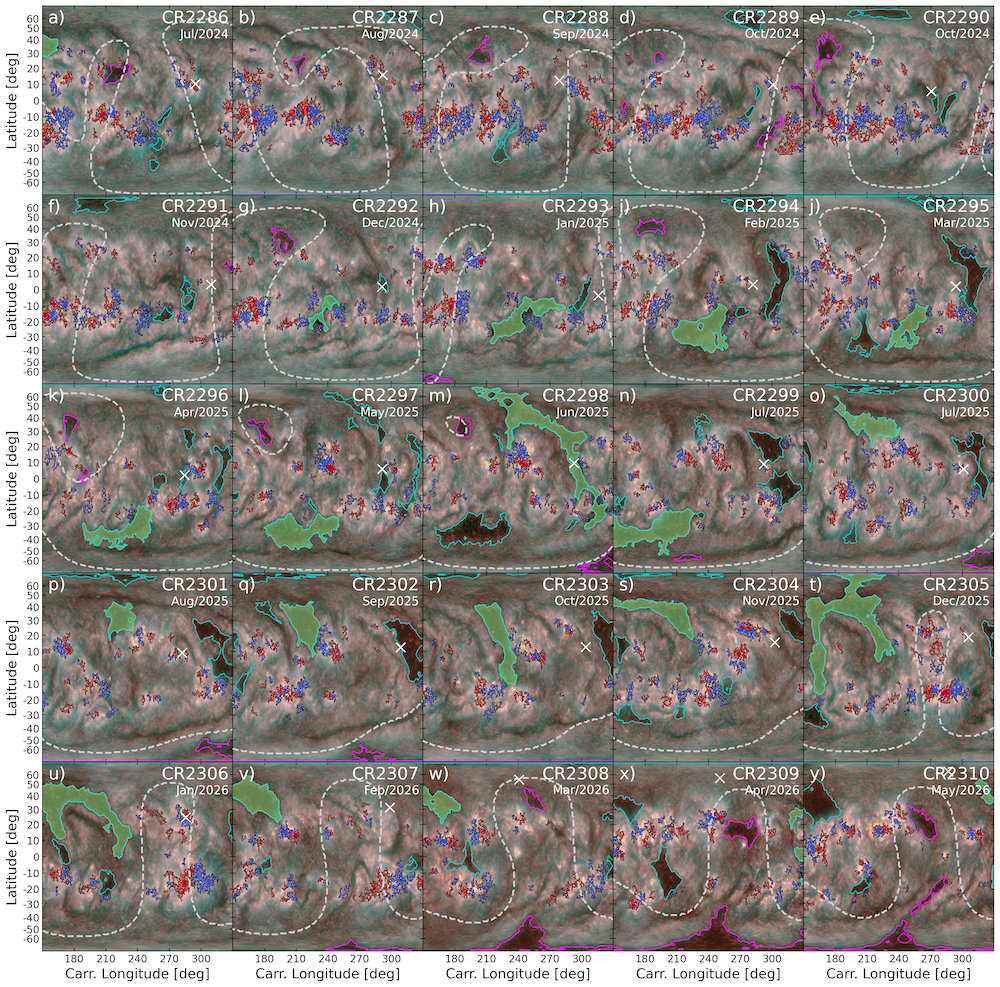}
    \caption{Same as Fig.~\ref{fig:coronalHoleSummary}, but now for coronal holes around the tilted-dipole's negative pole. The largest negative coronal hole is filled green.}
    \label{fig:coronalHoleSummary_negative}
\end{figure*}

In Sect.~\ref{sect:3:hole}, we tracked the evolution of persistent coronal holes around the tilted-dipole's positive pole from July 2024 to May 2026. Most of the positive coronal hole area was concentrated in a single large patch, covering up to $\sim8\%$ of the solar surface. However, negative coronal holes only covered up to $6\%$ of the solar surface; 25\% less than the maximum positive area. Figure~\ref{fig:coronalHoleSummary_negative} shows the coronal holes around the tilted-dipole's negative pole and highlights the largest negative coronal hole. In contrast to the positive pole, these Carrington longitudes were broken up by active regions that dominated July to November 2024 (panels a to f). Negative coronal holes began to develop in the southern hemisphere from unipolar magnetic regions forming at the leading edge of the AN1 and AN2 complex in December 2024 (panels g and h). Similar to the positive coronal hole area, the negative area first peaked in June 2025 (panel m) and then again in December 2025 (panel t). The second peak corresponded to the opening of the positive quadrupolar sector over the dipole's negative pole. Active regions that supported the growth of this positive sector were observed as early as May 2025 (panel l), with recurring flux emergence in that region continuing until January 2026 (panel u). Positive coronal holes began to form three months after the new sector opened (panel w). Comparing the top panels of Figs.~\ref{fig:coronalHoleSummary} and \ref{fig:coronalHoleSummary_negative}, negative coronal holes were far more fragmented, often split into four or more small patches, whereas positive coronal hole area was mostly contained in one large patch. We concluded that, flux emergence near the tilted-dipole's negative pole interrupted the growth of negative coronal holes, and triggered the formation of the new positive quadrupolar sector.

\end{appendix}
\end{document}